\documentclass[twocolumn,aps,pre,showpacs,amsmath,amssymb,floatfix]{revtex4-1}
\usepackage{graphicx}
\usepackage{dcolumn}
\usepackage{bm}

\begin{document}

\title{Collision Dynamics of Particle Clusters in a Two-dimensional Granular Gas}

\author{Justin C. Burton}
\email{Author to whom correspondence should be addressed:\newline justin.c.burton@emory.edu}
\author{Peter Y. Lu}
\author{Sidney R. Nagel}

\affiliation{James Franck Institute, Enrico Fermi Institute and Department of Physics, The University of Chicago}

\date{\today}

\begin{abstract} 
In a granular gas, inelastic collisions produce an instability in which the constituent particles cluster heterogeneously. These clusters then interact with each other, further decreasing their kinetic energy.  We report experiments of the free collisions of dense clusters of particles in a two-dimensional geometry. The particles are composed of solid CO$_{2}$, which float nearly frictionlessly on a hot surface due to sublimated vapor. After two dense clusters of $\approx$ 100 particles collide, there are two distinct stages of evolution.  First, the translational kinetic energy rapidly decreases by over 90\% as a ``jamming front" sweeps across each cluster. Subsequently, the kinetic energy decreases more slowly as the particles approach the container boundaries.  In this regime, the measured velocity distributions are non-Gaussian with long tails. Finally, we compare our experiments to computer simulations of colliding, two-dimensional, granular clusters composed of circular, viscoelastic particles with friction.  
\end{abstract}


\maketitle

\section{Introduction}

Collisions between inelastic particles produce a rich spectrum of nonequilibrium, many-body physics \cite{Jaeger1996,Kadanoff1999,Aranson2006,Goldhirsch2003}. Such interactions play a role in many natural phenomena such as the patterning of sand dunes \cite{Bagnold1954}, avalanche dynamics \cite{Jaeger1992}, the segregation of Saturn's dust rings \cite{Bridges1984,Bridges2001}, and even traffic patterns \cite{Helbing2001}. One well-studied example is the clustering instability that develops from an initially homogeneous granular gas of particles as it evolves in time \cite{Goldhirsch1993,McNamara1994,McNamara1996}. For elastic particles, energy is conserved and the component velocities in each spatial direction have a Gaussian distribution. However, when the particles are inelastic, the gas cools in an inhomogeneous manner so that particles cluster into dense areas where there is a higher frequency of collisions. The subsequent collisions between clusters become important for further evolution of the gas. 

Inelastic collisions are typically described by a coefficient of restitution, $\epsilon$, defined as the ratio of final to initial velocities upon normal impact. Much of the theoretical work in the realm of free granular gases has focused on the case of nearly elastic collisions (where $1-\epsilon \ll 1$) and on the initial stages of cooling where spatial density fluctuations are small and can be treated in a perturbative manner \cite{Brilliantov2004}. In addition, three-body collisions have been mostly ignored (i.e., the system is assumed to be dilute), so such theories cannot be applied to the internal dynamics of particles in a dense cluster. Alternatively, computer simulations of model granular gases have been successful in elucidating the role of a velocity-dependent coefficient of restitution \cite{Ramirez1999}, rotational degrees of freedom in the particles \cite{Luding1998,Brilliantov2007}, and arbitrary particle shapes \cite{Aspelmeier1998,Kanzaki2010}. Taken as a whole, recent studies suggest that the dynamics of granular gases are sensitive to the details of the particle interactions. Even the two most widely used computer algorithms have important discrepancies for oblique collisions \cite{Muller2012}.

The evolution of free granular systems pose an experimental challenge. In order to mitigate the influence of gravity or to measure dynamics over long times, most experiments are driven into a steady-state by vibration of the boundaries \cite{Olafsen1998,Rouyer2000,Losert1999,Harth2013}. A few experiments have been performed under microgravity conditions where particles are confined to a quasi-two-dimensional (2D) cell \cite{Tatsumi2009,Hou2008} for better visualization. However, it is known that collisions with the walls have a significant effect on the dynamics \cite{vanZon2004}. Recently, the initial stages of granular cooling have been measured using $\sim$ 50 diamagnetic particles trapped in a shallow potential, although the effects of a confining potential are  poorly understood \cite{Maass2008}. An elegant solution to many of these problems is to suspend the particles in a 2D layer so that collisions are confined to a plane. A recent example uses an air-table to suspend a dense gas of particles that are driven by collisions with the boundaries \cite{Nichol2012}.

\begin{figure}[!]
\begin{center}
\includegraphics[width=3.4 in]{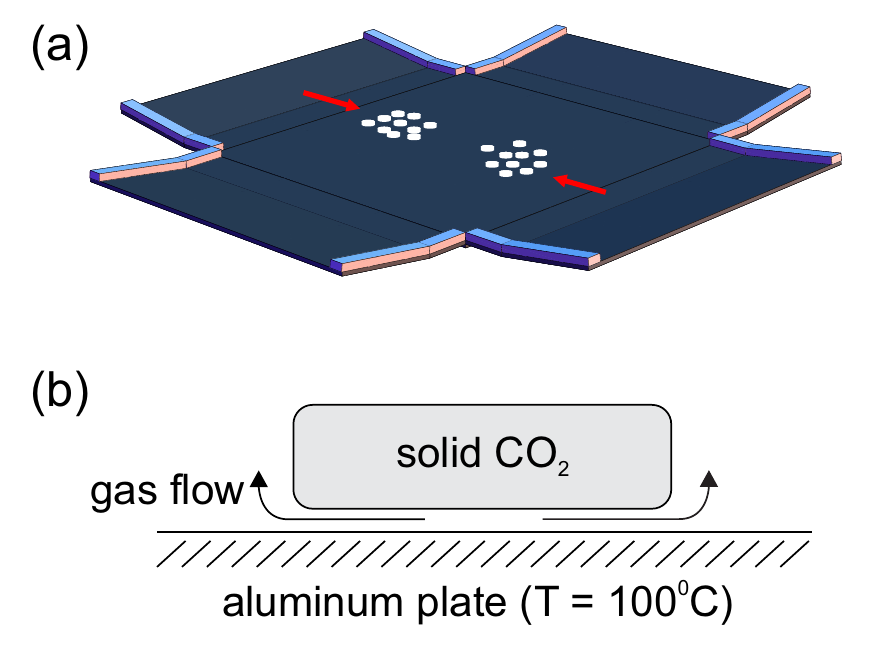}
\caption[]{(a) Schematic of the experimental apparatus. An anodized, aluminum plate with tilted boundaries is heated to $\approx$ 100$^\circ$C. Two clusters of solid CO$_{2}$ (dry ice) particles collide in the middle of the plate. Rectangular, silicone rubber strips prevent the particles from escaping from the plate edges. The dynamics are filmed from above with a high-speed camera. (b) Sublimated gas from beneath the dry-ice particles creates a high-pressure region which supports the particle's weight so that the particles float with nearly zero friction above the hot substrate. (Figure adapted from reference \cite{BurtonLu2013}).
} 
\label{platefigure}
\end{center} 
\end{figure}

In this paper, we focus on the collision of clusters of free granular particles, a process that contributes to the late-time evolution of a freely evolving granular gas. In this regime, the dynamics can involve highly dissipative collisions between particles that, depending on the particle size and surface interactions, may produce a complex set of outcomes such as sticking and partial fragmentation \cite{Ringl2012,Beitz2011}. Even ignoring such complexities, the cluster regime has been a difficult regime to study theoretically and simulationally as well as experimentally. Standard, near-equilibrium theories often lose accuracy in the highly dissipative regime \cite{Mitrano2012}. Early computer models of granular gases used constant coefficients of restitution for simplicity. Because of this, a numerical instability known as ``inelastic collapse" appeared for small values of $\epsilon$, where an infinite number of collisions occurred in finite time \cite{McNamara1996}. A similar singularity occurs in a hydrodynamic description of a granular gas, even in one dimension \cite{Efrati2005}. However, more realistic models of granular collisions resolved this instability by having $\epsilon$ depend on the impact velocity \cite{Brilliantov1996,Goldman1998}. In general, such issues underscore the necessity of laboratory experiments to guide our understanding of granular-gas dynamics.

In order to create freely-interacting granular particles in the laboratory, we use a two-dimensional system where levitation is achieved using the Leidenfrost effect, which is usually associated with liquid drops. When a drop of water is placed on a very hot pan, it will levitate on a thin, insulating cushion of evaporated vapor \cite{Burton2012,Quere2013}. Our experiments use a similar, recently reported, phenomenon involving the Leidenfrost levitation of solid particles \cite{Lagubeau2011}, where solid CO$_{2}$ (i.e., dry-ice) particles levitate on a layer of sublimated vapor. We characterize our particles by measuring the translational $E_{T}(t)$ and rotational $E_{R}(t)$ kinetic energy of two particles before and after an individual collision. We then investigate the collisions between two clusters of particles, each composed of $\approx$ 50--100 particles. By measuring the translational kinetic energy of the particles in the clusters, we find that there are two distinct regimes that occur after collision. Initially, there is a rapid decay of energy as a ``jamming front'' rapidly traverses each cluster \cite{BurtonLu2013}. Subsequently, the kinetic energy decays more gradually as the particles spread apart and collide with the container boundaries.  Single component velocity distributions, $P(v_x)$ and $P(v_y)$, (where $x$ and $y$ respectively label the direction of the initial velocity before collision and the transverse direction) measured after the initial collision are non-Gaussian.  They are sharply peaked near zero and have long tails.

We compare our results to computer simulations of two-dimensional, impacting granular clusters, each composed of circular, viscoelastic particles with frictional interparticle interactions. We directly simulate the experiment by using our measured initial positions and velocities as inputs to the simulations. Qualitatively, the agreement between experiment and simulation is excellent with a few quantitative differences that will be discussed. Thus the experiments provide a direct benchmark for the computer simulations.  The experimental and simulation methods will be described first in sections \ref{Experimental Methods} and \ref{Simulation Methods}, respectively, and then results will be discussed in section \ref{Results and Discussion}. 

\section{Experimental Methods}
\label{Experimental Methods}

The disk-shaped particles used in our experiments were cut from initially long, cylindrical pieces of solid CO$_2$ (dry ice). The pieces were obtained from Continental Carbonic Products Inc. Each particle had a radius of $\approx$ 0.8 cm and was $\approx$ 1.0 cm in height. To minimize water vapor condensation and unwanted sublimation, the particles were cut on-demand from fresh pieces of CO$_2$. However, due to the variability of the quality of dry ice on a daily basis, experiments were performed only when the dry ice had not sublimated significantly or condensed frozen water vapor. Usually the surface of the dry ice was rough so that diffuse reflection of light caused the particles to appear white in the video although occasionally it was very uniform and almost transparent so that it was more difficult to identify. 

A cast aluminum plate (MIC 6$\textregistered$, McMaster-Carr) of dimensions 61.0 cm $\times$ 61.0 cm $\times$ 1.25 cm, heated to $\approx$ 100$^\circ$C, was used to levitate the particles (Fig.\ \ref{platefigure}). Cast aluminum was necessary to maximize flatness of the material and to minimize thermal stresses when heated. Flatness was quoted at $\approx$ 380 $\mu$m over the length of the plate by the manufacturer although this was not directly measured. In order to heat the plate uniformly, a 61 cm $\times$ 61 cm flexible heater (maximum 1440 Watts) was attached to the back of the plate with thermal paste. Adhesive was not used so as to minimize stresses on the cast aluminum during heating and cooling. Finally, the apparatus was mounted on a larger, multipurpose aluminum plate (alloy 6061) with three leveling feet.  Thus the entire apparatus could be leveled so that the particles  experienced a two-dimensional, essentially force-free environment during the experiments. 

In order to provide ``reflecting", elastic boundary conditions for the particles at the edges of the plate, we initially tried various types of silicone rubber, which did not produce repeatable results. Instead, we choose to use a ramp at each edge to reflect the particles. Thus, attached to each edge of the cast aluminum plate was a thinner aluminum plate of dimensions 61 cm $\times$ 25 cm $\times$ 0.5 cm. Each of these four plates were bent at a slight angle ($\approx$ 8 degrees).  As the incident particles rose on the ramp, the gravitational potential energy stored and released the particles' kinetic energy.   We found that this procedure best mimicked elastic boundary conditions, where particles that approached the bend in the plates were specularly reflected by gravity. Silicone rubber walls were retained at the remaining edges of the plate to prevent particles from escaping the apparatus.

Each experiment was filmed from above using a high-speed digital camera (Phantom v9.0, Vision Research). The image size was 1200 pixels $\times$ 1200 pixels, which provided a resolution of 12.5 pixels/cm. All videos were recorded at 100 frames per second with an exposure time of 500 ms. To maximize contrast and to avoid spurious reflections from the aluminum surface, the entire apparatus was anodized black, so that the  dry-ice particles appeared white on a dark background in the videos. Most experiments consisted of the collision between two clusters of particles, each composed of $\approx$ 50--100 close-packed particles, which were deposited near the edges of the sloped boundaries and held in place with circular, plastic retainers. Upon release (manual removal of the retainer), the clusters gained momentum by sliding down the sloped boundaries, and subsequently collided near the middle of the plate. The initial speed of the particles upon impact was $\approx$ 50 cm/s. 

We tried several methods to measure quantities such as the kinetic energy during the collision process. Direct identification and tracking of each particle from frame-to-frame had many problems. Depending on the quality of the ice, the particles could be partially transparent and non-uniform in brightness, making them difficult to identify by computer algorithm. The non-circular shape of some particles also led to difficulties in identification when they were close packed, with adjacent flat edges in contact. Thus, we used a robust, particle-image-velocimetry (PIV) method, which correlates sub-sections of the image at frame $n$ with the image at frame $n+1$. The mass in each sub-section was measured by making the image binary, so that particles appeared white on a black background, and then counting the number of white pixels. This custom software was written in Mathematica 9.0 (Wolfram Research). We tested this method on images generated from a computer simulation of colliding granular clusters and found it to be reliable. However, our PIV software is only sensitive to translational motion, so particle rotations were not measured for cluster collisions.  As we show below, rotations contribute only an insignificant amount to the total kinetic energy of the particles, so that neglecting rotations does not adversely affect our results.

The source of levitation for the particles is sublimated gas flow due to the heat from the aluminum plate, which caused the particles to lose mass during the experiments. We found that over the course of one 30 second experiment, the particles loss approximately 40\% of their mass. This occurred essentially in the vertical direction, so that the height of the particles decreased, but their shape and size did not change when viewed from above. This is an important feature for our analysis. We assume that all of the particles lose mass uniformly, so that the trajectories of the particles after a collision only depends on the $\textit{ratio}$ of masses, not the absolute value. Thus, even though the particles are continually losing mass, the positions and velocities of the particles are not sensitive to this mass loss. It should be noted that throughout the remainder of the discussion, the kinetic energies that we measure are from our two-dimensional images and do not take this mass loss into account.  In addition, because the propagation of the jamming front, which is the first regime of the dynamics after a cluster collision, is very rapid and occurs in much less than 1 second, the entire issue of mass loss is insignificant in that regime.

\section{Simulation Methods}
\label{Simulation Methods}

Our simulations use two-dimensional, time-integrated molecular dynamics to model granular particles \cite{Poschel2005}. All particles are monodisperse circles with radius, $\sigma$, and are confined to a square box with elastic boundary conditions (i.e. specular reflection upon impact). The particles have two translational degrees of freedom ($x$ and $y$ position), as well as one rotational degree of freedom (angle $\theta$), which corresponds to rotation about the center of mass of each circle. When two particles overlap, they experience both elastic and dissipative forces. First, the particles experience a finite-ranged, Hertzian repulsive force due to the interaction potential \cite{OHern2003}:
\begin{align} 
V(r_{ij})= 
\begin{cases} 
\frac{2V_{0}}{5}\left(1-\frac{r_{ij}}{2\sigma}\right)^{5/2}
 & \text{$r_{ij}<2\sigma$,}
\\
0 &\text{$r_{ij}>2\sigma$,}
\end{cases}
\label{int_potential} 
\end{align}  
where $V_{0}$ sets the energy scale of the interaction, $r_{ij}$ is the distance between the centers of the particles $i$ and $j$, and the elastic force on particle $i$ is $\vec{F}_{e}=-\vec{\nabla} V$, where the gradient is taken with respect to the coordinates $\vec{r}_i$.  All lengths in our simulations are scaled by the particle size $\sigma$, all masses by the particle mass $m$, and all times by the time scale $\sigma\sqrt{m/V_{0}}$. 

Although our CO$_{2}$ particles collide in two dimensions, here we use a three-dimensional, Hertzian contact force law (exponent = 5/2) because particle contact likely occurs at a small, point-like regime near an asperity at the particle surface.  Subject to this force alone, energy would be conserved during the collision process. Thus, the particles also experience a viscous component to the normal force \cite{Brilliantov1996}:
\begin{align} 
\vec{F}_{v}=A\frac{d\xi}{dt}\frac{d\vec{F}_{e}}{d\xi},
\label{viscous_force} 
\end{align}  
where the overlap $\xi=2\sigma-r_{ij}$. The prefactor $A$ represents the viscous relaxation time scale, and should be less than the interaction time. Both $F_{e}$ and $F_{v}$ are normal to the surface at the contact zone, thus, rotational motion cannot be induced using these forces. 

We also include tangential friction in the simplest way possible. At large velocities, the frictional force $F_{f}$ should be proportional to the total normal force $F_n=F_e+F_v$, with a coefficient of friction $\mu$. At lower velocities, this force should approach zero in a smooth way, since discontinuities can cause numerical instabilities. The most common way to do this is in a piecewise fashion:
\begin{align} 
\vec{F}_{f}=-\mathrm{sgn}(v_t)\times \min\left\{\beta v_t,\mu|F_n|\right\}\hat{t},
\label{friction_force} 
\end{align}  
where $\hat{t}$ is a unit vector pointing tangential to the surface at the point of contact. The relative tangential velocity $v_t$ between the particles at the point of contact will depend on the translational velocities and angular velocities, and the constant $\beta$ sets the velocity scale for when Coulombic sliding friction sets in (i.e., $F_f=\mu F_n$).

For all simulations, time steps were chosen so that energy was conserved and the numerical scheme was stable when the collisions lacked dissipation ($A$ = 0, $\mu$ = 0). Since we are using molecular dynamics, restitution coefficients (normal $\epsilon_n$ and tangential $\epsilon_t$) do not directly enter as parameters. During a collision, the resulting loss of kinetic energy will depend on the values of $A$, $\beta$, and $\mu$. For all our simulations, we use the values $A$ = 10.0, $\beta$ = 1.0, and $\mu$ = 0.5, which were chosen to best match the kinetic-energy loss measured in experiments. In addition, the velocity scale $c=\sqrt{V_{0}/m}$ characterizes the speed of sound in our simulations. Although the speed of sound in different ices can vary considerably depending on the crystallinity and porosity, we use a characteristic value of $c$ = 300,000 cm/s when directly simulating the experimental data.

\section{Results and Discussion}
\label{Results and Discussion}

\subsection{Individual Particle Collisions}

\begin{figure}[!]
\begin{center}
\includegraphics[width=3.3 in]{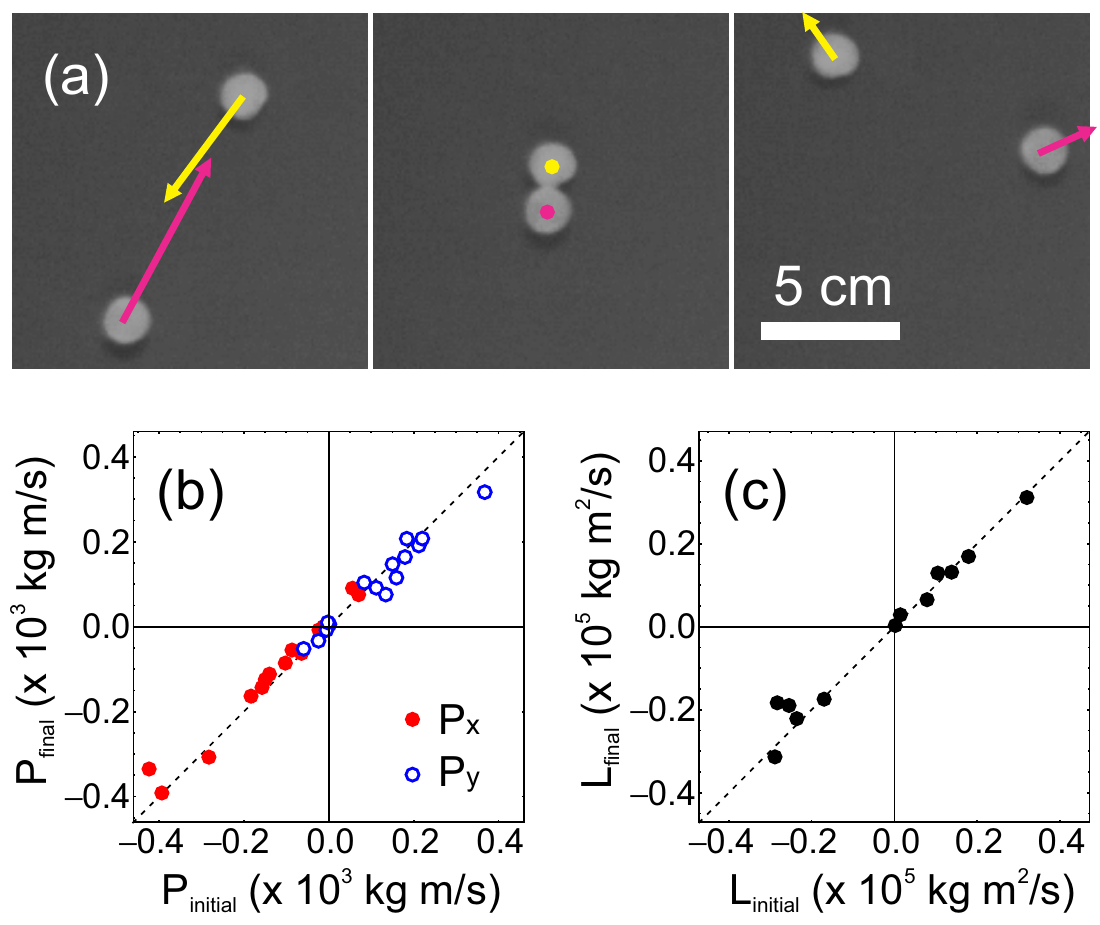}
\caption[]{(a) Images showing the collision of two isolated CO$_{2}$ particles. The collision induces changes in each particle's velocity and rotation. The time between the first and last image is 0.85 seconds. The arrows indicate the direction and relative magnitude of the velocity. (b) Final momentum versus initial momentum, and (c) final angular momentum versus initial angular momentum of both particles for 15 separate collisions. The dotted lines indicate perfect conservation of translational and angular momentum.
} 
\label{conservation}
\end{center} 
\end{figure}

In order to characterize the collisions between dry-ice particles, we measured the masses, moments of inertia, and translational and rotational velocities of individual particles for 15 independent collisions. This was done by single particle tracking.  Fig.\ \ref{conservation}a shows typical frames from a movie of a single collision. We investigated collisions that occurred with relative velocities between 30 and 80 cm/s. An important feature of our experiment is that the particles float above the aluminum surface and undergo free collisions. During an impact event, if the particles touch the surface, then the momentum that we measure will not be conserved and kinetic energy will be lost due to friction with the plate. 

To quantify this effect, we plot the final momentum and angular momentum versus initial momentum  in Figs.\ \ref{conservation}b and 2c, respectively. In general, momentum is conserved even when particles are colliding at speeds up to 80 cm/s, although there is a slight spread in the data points. Larger excursions from ideal momentum conservation (shown by the dotted lines) are likely due to a transient contact with the aluminum plate. In addition, multi-body collisions could produce larger forces that may cause particles to buckle out-of-plane and touch the aluminum. As discussed below, there is some evidence for this in our analysis of the collisions of particle clusters. 

For ordinary (H$_2$O) ice particles, more kinetic energy is lost at higher impact speeds \cite{Bridges1984}. In reference \cite{BurtonLu2013}, we showed that, for our dry-ice particles, the ratio of the total final kinetic energy to total initial kinetic energy varies considerably from one collision event to another and typically falls in the range of $E_{final}/E_{initial}$ = 0.1--0.6. The energy loss will depend strongly on the impact parameter of the collisions, which is the most likely reason for the large spread in this value. For normal, head-on collisions with negligible rotation energy, the energy loss can be related to the normal coefficient of restitution: $E_{final}/E_{initial}\approx\epsilon_n^2$. This is an upper bound on the measured value of $\epsilon_n$. Even the most ``elastic" collisions between dry-ice particles result in a loss of $\approx$ 40\% of the kinetic energy, corresponding to $\epsilon_n\approx0.78$. 

\begin{figure}[!]
\begin{center}
\includegraphics[width=3.3 in]{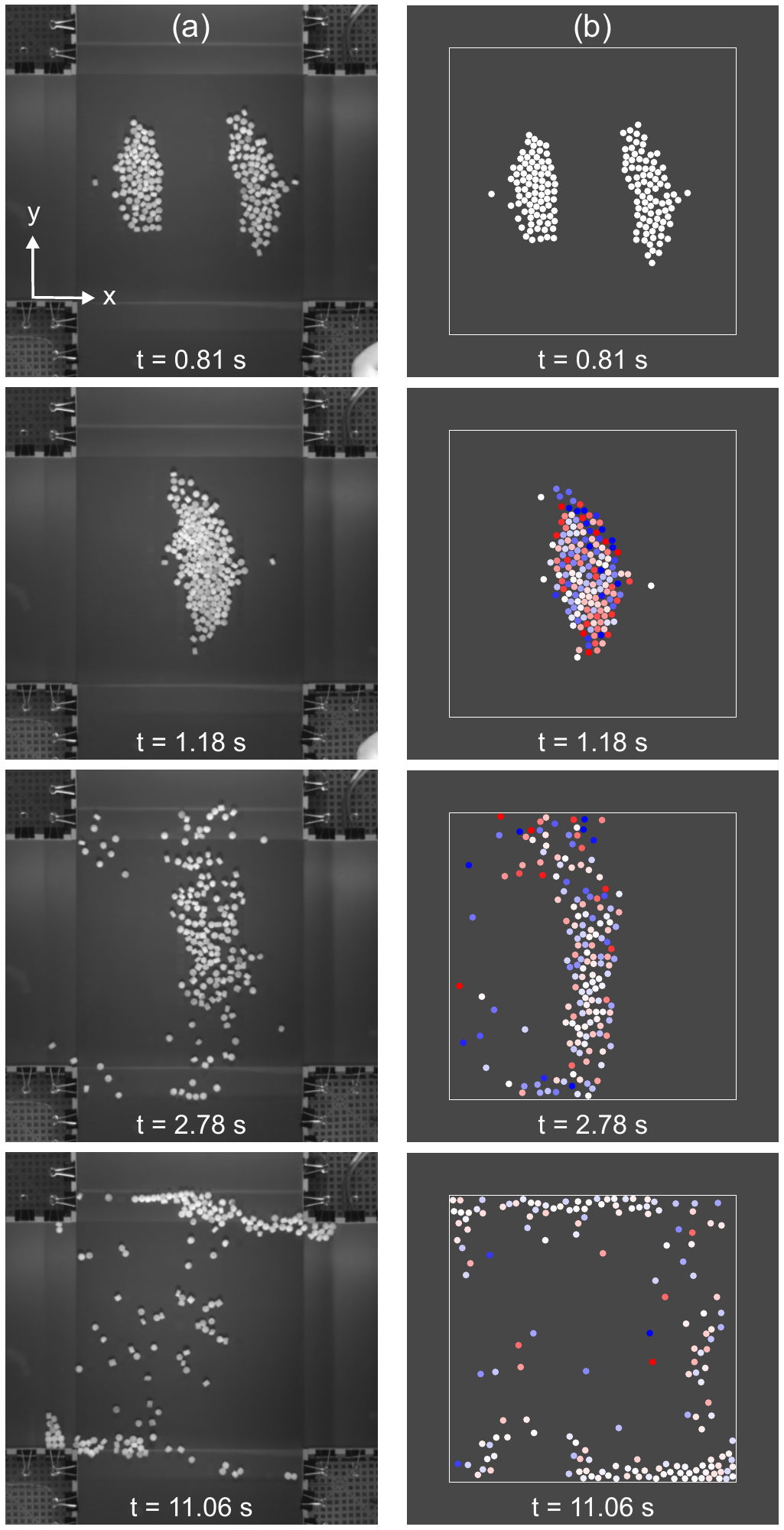}
\caption[]{(a) Images from a typical movie showing the collision in the $x$-direction of two clusters after sliding down the sloped edges of the apparatus. The particles spread out in the $y$-direction, and eventually come to rest in clusters near opposite edges of the plate. (b) Simulation of the same experiment using viscoelastic particles with friction. The color corresponds to the magnitude of angular velocity $\omega$ in the clockwise (red) and counterclockwise (blue) direction, where white indicates $\omega$ = 0. Clustering also occurs near the boundaries in the simulation.
} 
\label{dry_ice_1}
\end{center} 
\end{figure}

\subsection{Cluster Collisions}

\begin{figure}[!]
\begin{center}
\includegraphics[width=3.3 in]{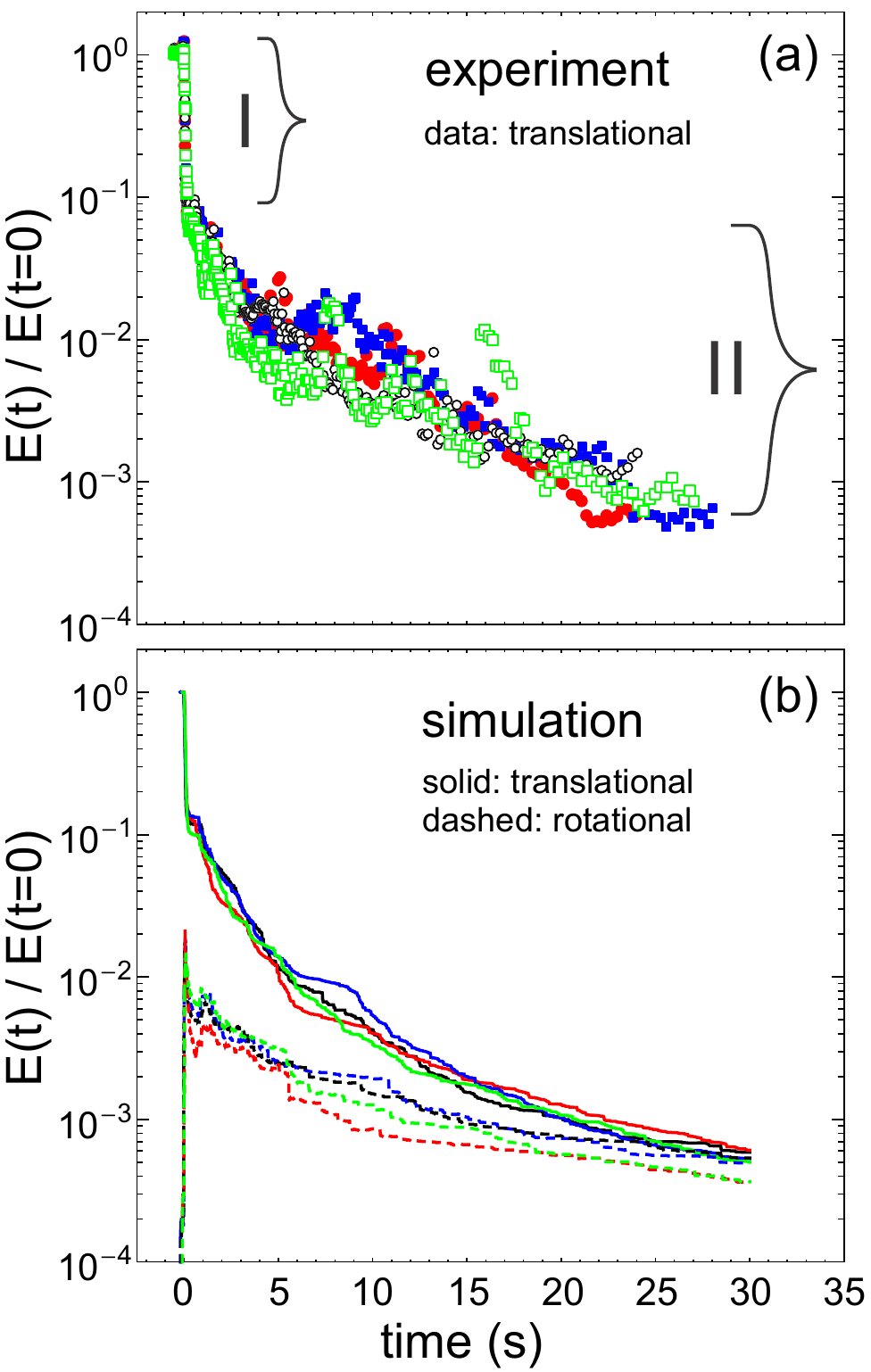}
\caption[]{(a) Translational kinetic energy $E_{T}(t)$ for four different experiments. The first particles collide at $t$ = 0.  Each data set is normalized by the initial kinetic energy prior to collision, $E(t=0)$. The initial drop in kinetic energy (regime I) is discussed in section \ref{jam_front}, and the late-time decay (regime II) is discussed in section \ref{late_time}. (b) $E_{T}(t)$ (solid) and $E_{R}(t)$ (dashed) from the direct simulations of the experiments, both normalized by $E(t=0)$. Colors refer to the corresponding experimental data set in (a).
} 
\label{all_energy}
\end{center} 
\end{figure}

\begin{figure*}[!]
\begin{center}
\includegraphics[width=6.6 in]{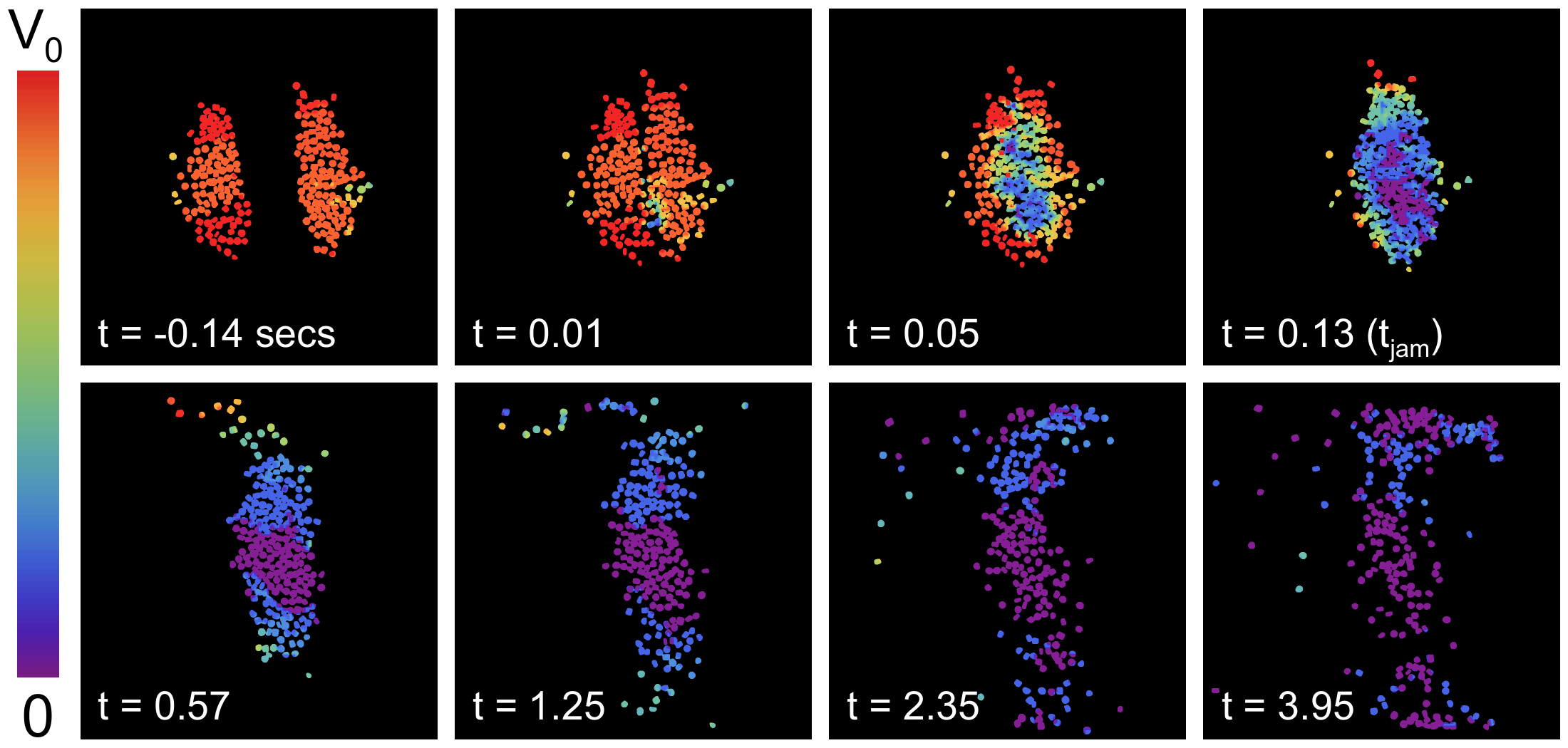}
\caption[]{Images of the particles from an experiment during the initial moments of the collision. The color indicates the magnitude of the velocity, as denoted by the scale on the left ($v_{0}$ = 52 cm/s). The front of reduced velocity (jammed particles) travels faster than the initial speed of the particles and eventually encompasses all of the particles.  The cluster then proceeds to elongate in the $y$-direction until it touches the boundary of the system, inducing further collisions among the particles.
} 
\label{velocity_color}
\end{center} 
\end{figure*}

Next, we examine the impact of two clusters, each composed of $\approx$ 50--100 CO$_{2}$ particles. Fig.\ \ref{dry_ice_1}a shows four frames from a typical video. Although the clusters are initially released in a semi-circular shape, they are elongated because particles near the rear travel farther down the sloped edges, and thus gain more energy. After impact, the resulting single cluster elongates in the $y$-direction, with particles impacting the top and bottom boundaries of the container. Many collisions occur near the boundaries, so the particles tend to cluster in these regions, where they remain until the end of the experiment. Although the aluminum plate is quite flat and the apparatus is leveled before the experiment, the particles tended to move towards opposite corners over long times. Without friction from the plate, even the smallest gravitational bias can be observed.

Fig.\ \ref{dry_ice_1}b shows a simulation of the experiment. Initial particle positions and velocities were measured as inputs to the simulation. The colors represent the angular velocity of the particles. Initially, the particles have zero rotation, but upon impact, particles begin to rotate. Near the end of the simulation, particles have clustered mostly near the boundaries, just as in the experiment. 

\subsection{Kinetic Energy}

A more quantitative comparison between the experiment and simulation comes from the time evolution of the kinetic energy. Fig.\ \ref{all_energy}a shows the translational kinetic energy $E_{T}(t)$ from four independent experiments, normalized by $E(t=0)$. There are two main regimes shown in the data. After the impact of the clusters at $t$ = 0 s, there is an initial, sharp drop in the translational energy.  Subsequently, there is a second regime where the energy slowly decays as the large jammed cluster expands. We will separately examine these two regimes in more detail. 

We quantitatively compare the data to direct computer simulations of the four experiments. In Fig.\ \ref{all_energy}b, we plot $E_{T}(t)$ from the simulations, which show the same qualitative features as in the experiment. In addition, we also plot the rotational component, $E_{R}(t)$, for each simulation.  These simulations show that $E_{R}(t)$ constitutes only a very small fraction of the total energy until late times when most of the particles have clustered in the corners.  The few particles remaining near the center have residual rotational energy which does not decay since in that region there are no further collisions.  

During the initial impact, the experiment shows a larger drop in $E_{T}$ than does the simulation. We found that even if the particles in the simulation were perfectly inelastic, this initial drop could not be matched in magnitude. We suspect that some particles in the experiment, upon initial impact, buckle out-of-plane and touch the aluminum plate. This would contribute a source of energy loss not considered in the simulations.

\subsection{Region I: Propagation of a Jamming Front}
\label{jam_front}

In order to explore the initial regime of the cluster collisions in more detail, we look at the spatial distribution of particle velocities.
Fig.\ \ref{velocity_color} shows images from one of the experiments in which each particle has been colored according to its velocity magnitude obtained from the PIV analysis. The top row of images lie in the first regime ($0<t<t_{jam}$), where a ``jamming front'' sweeps across each cluster and rapidly dissipates kinetic energy. The time $t_{jam}$ denotes when the jamming front has encompassed all of the particles. The bottom row of images lie in the late-time regime ($t>t_{jam}$). 

\begin{figure}[!]
\begin{center}
\includegraphics[width=3.2 in]{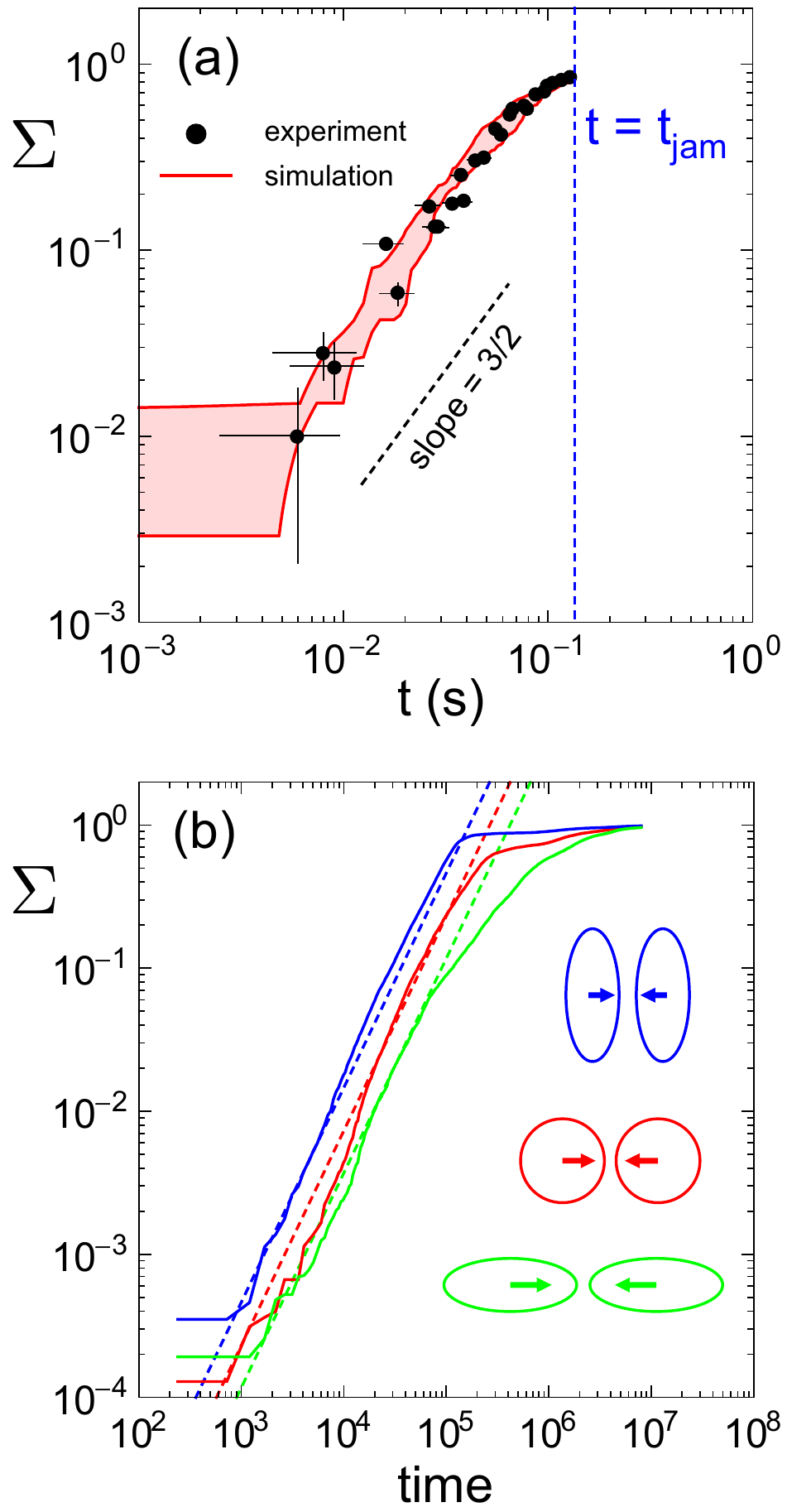}
\caption[]{(a) $\Sigma$ versus time averaged over 4 different experiments from Fig.\ \ref{all_energy}. The error bars (experiment) represent one standard deviation from the mean, and the solid red area shows the maximum and minimum values from the simulations. At early times the data is flat because only two particles have collided. Both the data and simulations are consistent with $\Sigma\propto t^{3/2}$. (b) $\Sigma$ versus time from simulations of elliptical-cluster collisions with 5000 particles per cluster. The cluster shapes are shown on the right with the corresponding colors. The dashed lines are predictions from eqn.\ \ref{energyrat} with parameters $\phi_J$ = 0.907, $\phi_{0}$ = 0.71, $v_0$ = 1.6 $\times$ 10$^{-4}$, and $a$ = \{52.7, 83.7, 131.8\} (blue, red, green). 
} 
\label{1st_regime}
\end{center} 
\end{figure}

This jamming front has previously been reported \cite{BurtonLu2013}.  At $t=0$, a region of reduced velocity appears as the colliding particles suddenly lose their forward momentum. As more particles collide from the rear of each cluster, the jammed region grows. This jamming front is similar to dynamic jamming fronts in suspensions \cite{Waitukaitis2012,Waitukaitis2013}. The front speed depends on the area fraction, $\phi_{0}$, of the particles. For high densities, close to the theoretical maximum for two-dimensional, monodisperse packings ($\phi_{J}=\pi$/$\sqrt{12}$ $\approx$ 0.907), the front would travel close to the speed of sound in a single particle \cite{Gomez2012}. As the density is reduced, it takes longer for the particles near the rear of the cluster to impact their neighbors, so that the kinetic energy will also decay more slowly. 

We compare the decay in kinetic energy of the simulations with that found in the experiments. In Fig.\ \ref{1st_regime}a, we plot the normalized energy loss, $\Sigma(t) \equiv (E(t=0)-E_{T}(t))/E(t=0)$, versus $\tau \equiv t/t_{jam}$ on a log-log plot. The experimental data (black points) are limited by the time resolution of the video (100 frames per second). In this range, the simulations (red lines) agree well with the data. At times close to $t=0$, the simulation data is flat because only two particles have collided. Eventually they collide with their neighbors and the jamming front spreads causing a further reduction in kinetic energy. 

We also simulate the collision of two elliptical clusters, each composed of 5000 particles. Fig.\ \ref{1st_regime}b shows $\Sigma$ versus time for the initial moments after impact. The blue line represents ellipses with an aspect ratio of 5/2 \cite{BurtonLu2013}, the red line represents perfect circular clusters, and the green line represents an aspect ratio of 2/5. The area fraction of particles inside the clusters for all three simulations is $\phi_{0}$ = 0.71. Although the prefactor is affected by the initial cluster geometry, all three show power-law behavior with an exponent consistent with 3/2. 

This power-law behavior can be explained by the geometry and density of the clusters. If all of the particles in the jammed region have lost their kinetic energy, then $\Sigma$ is predicted to be \cite{BurtonLu2013}:
\begin{align} 
\Sigma(t)=\frac{4 \sqrt{2}}{3 \pi}\left(\frac{\phi_J}{\phi_J-\phi_0}\right)\left(\frac{v_0 t}{a}\right)^{3/2}.
\label{energyrat} 
\end{align}  
The constant $a$ is the length of the axis of the ellipse in the direction of motion. This prediction is shown by the dashed lines in Fig.\ \ref{1st_regime}b for each different cluster geometry. Although here we have varied the geometry of the clusters, our results from reference \cite{BurtonLu2013} shows that the energy decay is virturally independent of the particles' normal and tangential restiution, in agreement with the model. This is because particles lose nearly all of their kinetic energy as the jamming front propagates, which can occur through a few highly inelastic collisions, or many weak inelastic collisions. 

\subsection{Regime II: Late-time decay}
\label{late_time}

\begin{figure}[!]
\begin{center}
\includegraphics[width=3.2 in]{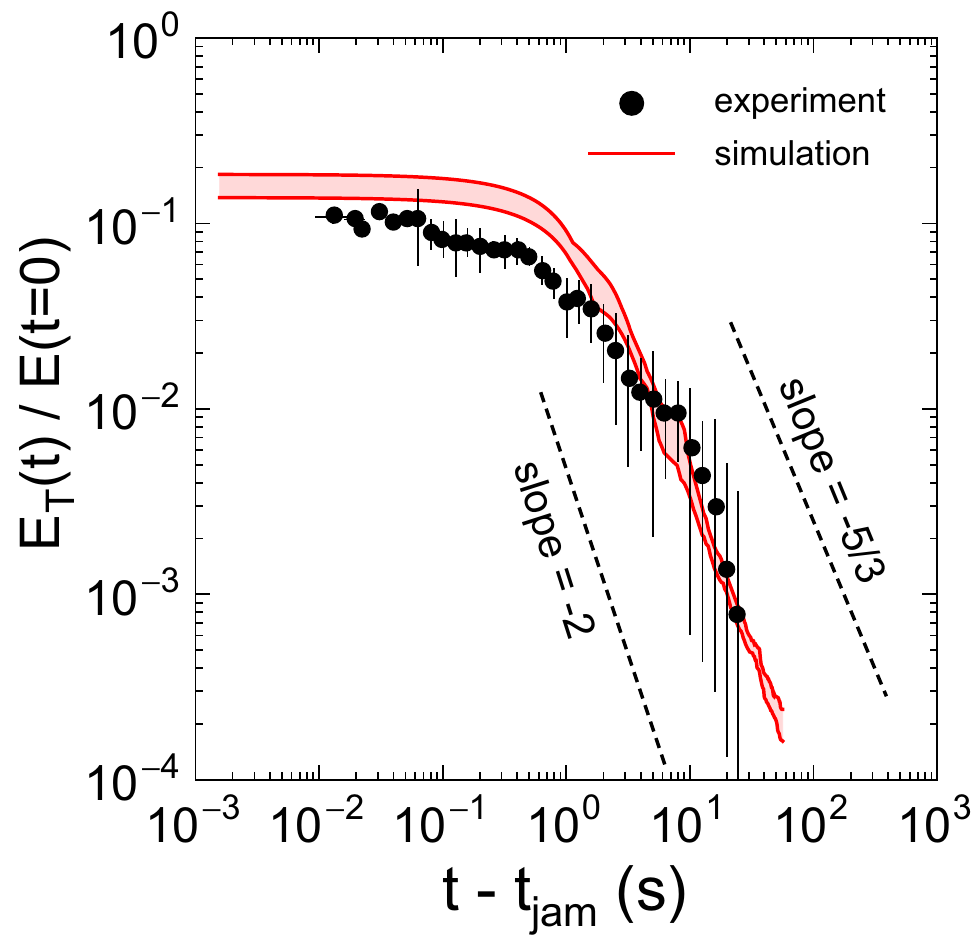}
\caption[]{(a) Translational kinetic energy $E_{T}(t)$ for four different experiments. Each data set is normalized by the initial kinetic energy prior to collision, $E(t=0)$. Just before $t$ $\approx$ 1.0 s, the particles hit the boundaries of the plate and begin a slow decay of the kinetic energy due to multiple collisions. The error bars (experiment) represent one standard deviation from the mean, and the solid red area shows the maximum and minimum values from the simulations. At late times, both experiments and simulations are consistent with $E_{T}(t)/E(t=0)\propto (t-t_{jam})^{1.8\pm0.2}$.} 
\label{2nd_regime}
\end{center} 
\end{figure}

After the jamming front has propagated through the system, the cluster continues to expand in the $y$-direction. In this regime, we find slower decay of the translational energy.  We show the data for $E_{T}(t)$ versus $t-t_{jam}$ in Fig.\ \ref{2nd_regime}. In many dynamical regimes of granular gases, the kinetic energy decays in a power law fashion ($E\propto t^{\gamma}$). Taking the experiment and simulation data as a whole, our results suggest that the energy decays in time with an exponent $\gamma=1.8\pm0.2$. This range encompasses the expected value for a gas of particles with constant $\epsilon$ in the homogeneous cooling regime (density is approximately uniform), where $\gamma$ = 2 (Haff's law), as well as the expected value for a gas of viscoelastic particles, where $\gamma=5/3$ \cite{Brilliantov2000}. However, for highly dissipative particles such as ours, others have obtained exponents closer to $\gamma$ = 1.2 for homogeneous cooling \cite{Kanzaki2010}. Our measurements in this late-time, ``slow decay", regime best represent a system cooling by particles collecting into clusters at the boundaries. In a completely free system or one with periodic boundary conditions, this regime may be transient as the clusters may fragment, expand, or collide with other clusters.  

\begin{figure*}[!]
\begin{center}
\includegraphics[width=7.0 in]{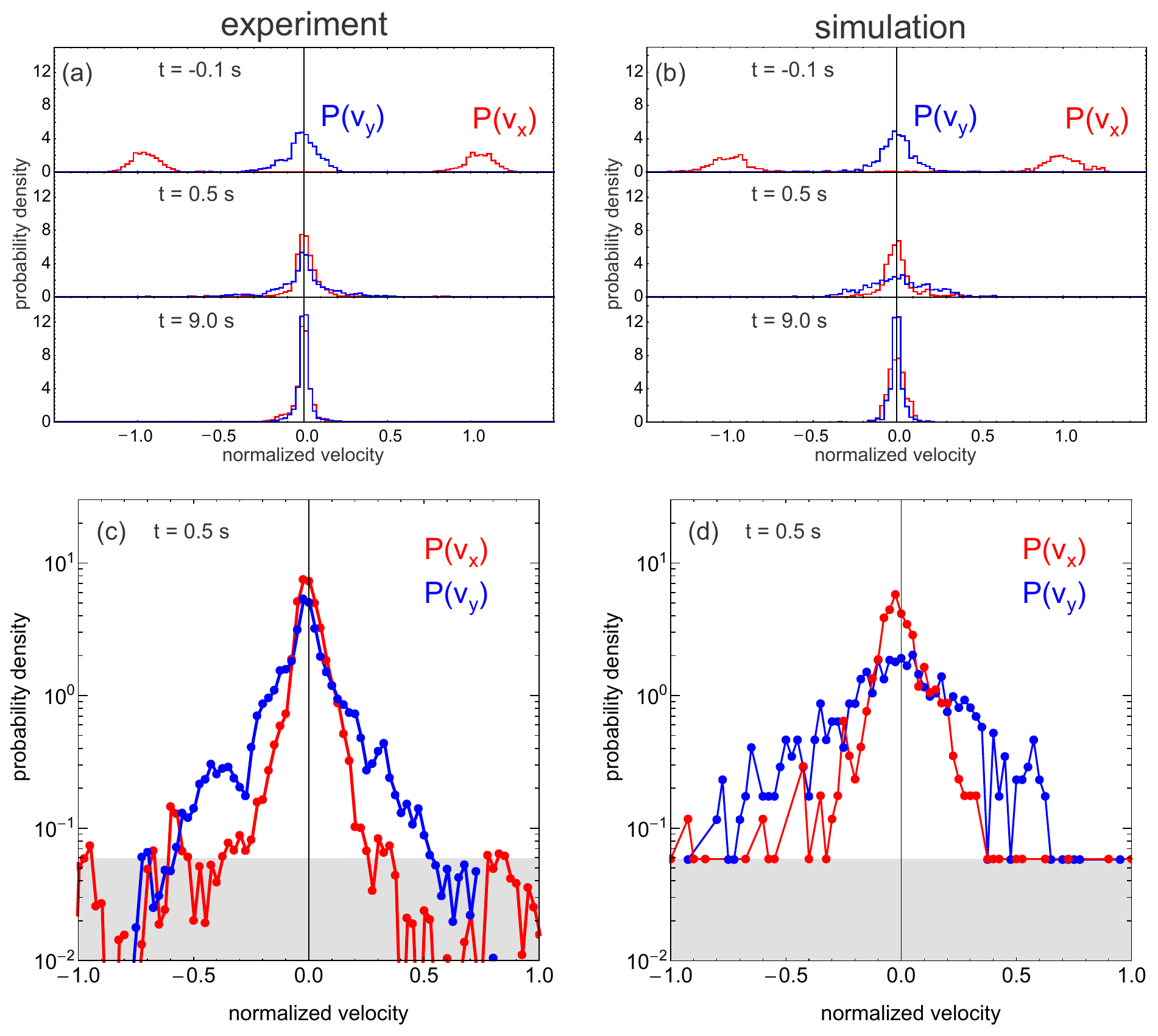}
\caption[]{Single component velocity probability distributions ($P(v_x)$ and $P(v_y)$) for both the experiment (a) and corresponding simulation (b). The distributions are averaged over four experiments. The clusters impact at $t$ = 0 s. All velocities are normalized by $v_0$. (c-d) Log-linear plot of $P(v_x)$ and $P(v_y)$ at $t$ = 0.5 s after the initial impact. Data in the light gray region (below density = 6 $\times$ 10$^{-2}$) is not reliable, and represents the velocity of a single particle in one bin size. Since particles are not directly tracked in the experiment, this limit shows up as an increase in noise.
} 
\label{vdists}
\end{center} 
\end{figure*}

During the collision of two clusters, an important feature which will determine further evolution of the granular particles is the probability distribution of translational velocities. We can also compare our experiment and simulation data by measuring the single component velocity distributions, $P(v_x)$ and $P(v_y)$. Fig.\ \ref{vdists}a shows the probability density for both the $x$ and $y$ direction, averaged over four experiments, at three different times. All velocities are normalized by the average initial velocity of the clusters in the $x$ direction ($v_{0}$). Before the collision, there are two peaks in $P(v_x)$ located near 1 and -1, which correspond to each cluster moving towards each other in opposite directions.  Ideally, if every particle had the same speed, these peaks would resemble Dirac-delta functions. In addition, there is some residual velocity in the $y$ direction, as shown by $P(v_y)$, which is centered around zero.

At $t$ = 0.5 s, just after the initial collision, both distributions become broad with a sharp peak in the center.  The particles in the middle of the cluster remain fairly stagnant with little or no velocity while particles near the edges of the newly-formed cluster move towards the boundaries. At $t$ = 9.0 s, most of the particles have drifted towards the boundaries and have little motion, and both $P(v_x)$ and $P(v_y)$ look similar. The tails of the distributions are dominated by a few particles near the center of the plate which are still moving. Fig.\ \ref{vdists}b shows the data for the four simulations that correspond to these experiments. For the most part, the velocity distributions from the simulation compare well with the experiment. However, there is slightly more clustering at the boundaries in the experiment due to the gravitational bias. This acts to equilibrate the $x$ and $y$ velocities through more collisions.

In Fig.\ \ref{vdists}c-d, we plot the velocity distributions at $t$ = 0.5 s on a log-linear plot for both the experiment (c) and the simulation (d). Data below a probability density $\approx$ 6 $\times$ 10$^{-2}$ is not reliable because this corresponds to a single particle. This limit can be seen in the simulation as a horizontal line of points, and as an increase in noise in the experiment. There is a sharp peak near zero velocity in both the experiment and simulation, and broad tails at higher velocities. This is somewhat more clearly seen in the experimental data. Although the data are quite noisy, the shapes of the distributions are consistent with an exponential form, i.e., a straight line on a log-linear plot. Exponential velocity distributions have also been measured for a freely cooling, dense gas of glass beads on a plastic substrate \cite{Losert1999}, as well as in hydrodynamic calculations \cite{Esipov1997}.

\section{Conclusions}

Our experiments and complementary simulations provide a detailed picture of the time evolution of the particle dynamics after the impact between two dense, granular clusters. Such clusters are a generic feature in freely-evolving granular systems.  We find that a large fraction of the system's kinetic energy is lost over a short time scale as a jamming front traverses the particles in each cluster. The energy decays as a power law in time.   This result relies on the particles in a jammed region losing much of its kinetic energy.  This would occur when, as in our experiments with CO$_2$ particles, the coefficient of restitution is small.  We speculate that this result may also be valid for larger coefficients of restitution if the number of particles are also increased.  Since most granular particles are rather inelastic except at very small velocities,  we expect our results to be applicable to wide variety of situations including astrophysical phenomena whenever particle clusters collide. 

Our simulations were matched to our experiments so that they started with the same the initial positions and velocities of the particles in the experiment as well as with similar restitution coefficients.  Thus, they generally provide an excellent benchmark for granular simulations.  We can conclude that typical molecular dynamics simulation methods using circular particles agree rather well with the experimental data. This is encouraging, since the particle number in some of the simulations was rather small ($<$ 200), so that the system behavior cannot be obtained from statistical averaging. In addition, although we did not track the rotation of our particles, the simulations suggest that during most of the evolution of the collision process, the rotational kinetic energy is negligible.

After the initial head-on collision between granular clusters, the kinetic energy is reduced to $\approx$ 5--20\% of its initial value. This seems to be independent of the number of particles, as we see it in experiments and simulations with $\approx$ 100 particles per cluster, and simulations with 5000 particles per cluster. This number is also independent of the inelasticity of the particles, so long as they are sufficiently inelastic. We also observe that single-component velocity distributions obtained from the experiments suggest an exponential form, which is consistent with previous freely-evolving experiments and calculations of dense granular systems in the literature, albeit with different geometries and initial conditions. 

\section{Acknowledgements}

We are grateful to Helmut Krebs for invaluable advice and assistance.  We thank Efi Efrati, Narayanan Menon and Scott Waitukaitis for important discussions. We acknowledge support from the Illinois Math and Science Academy (P.Y.L.), NSF MRSEC DMR-0820054 and PREM DMR-0934192 (J.C.B.) and the US Department of Energy, Office of Basic Energy Sciences, Division of Materials Sciences and Engineering, Award No. DE-FG02-03ER46088 (S.R.N.).

\end{document}